\documentclass[twocolumn,showpacs,superscriptaddress,prb]{revtex4}    
\usepackage[T1]{fontenc}
\usepackage{graphicx,amsmath,mathrsfs}
\usepackage{amssymb,color}
\usepackage{amsthm}
\usepackage{times}

\newcommand{\bs}[1]{\boldsymbol{#1}}

\def\ie{\emph{i.e.},\ }
\def\eg{\emph{e.g.}\ }

\newcommand{\pd}{{\phantom{\dag}}}
\newcommand{\up}{\uparrow}
\newcommand{\dw}{\downarrow}
\newcommand{\eps}{\varepsilon}

\begin{document}
\title{Giant Magneto-Resistance and Perfect Spin Filter in Silicene, Germanene, and Stanene}
\author{Stephan Rachel}
\affiliation{Institute for Theoretical Physics, TU Dresden, 01062 Dresden, Germany}
\author{Motohiko Ezawa}
\affiliation{Department of Applied Physics, University of Tokyo, Hongo 7-3-1, 113-8656, Japan}

\begin{abstract}
Silicene, Germanene and Stanene are two-dimensional topological insulators exhibiting helical edge states. We investigate global and local manipulations at the edges by exposing them to (i) a charge-density-wave order, (ii) a superconductor, (iii) an out-of-plane antiferromagnetic, and (iv) an in-plane antiferromagnetic  field. We show that these perturbations affect the helical edge states in a different fashion. As a consequence one can realize quantum spin-Hall effect without edge states. In addition, these edge manipulations lead to very promising applications: a giant magneto-resistance and a perfect spin-filter. We also investigate the effect of manipulations on a very few edge-sites of a topological insulator nanodisk.
\end{abstract}



\maketitle

\section{Introduction}
Topological phases play a major role in modern condensed matter physics. It is expected that topological effects will soon become more important for applications and technological developments. In general, topological phases are characterized by  non-local ``quantum numbers'' such as topological invariants\,\cite{thouless-82prl405,Volovik03,nijs-89prb4709,kane-05prl146802} or topological entanglement-entropy\,\cite{kitaev-06prl110404,levin-06prl110405}. This contrasts the paradigm of conventional symmetry broken phases, \eg magnets and superfluids, described by a local order parameter.
As a consequence of this non-locality, it is often said that microscopic details as well as local perturbations do not matter for topological phases.

There exist at least two types of topological orders, \textit{intrinsic topological order}\,\cite{wen90ijmpb239} and
\textit{symmetry protected topological order}\,%
\cite{Volovik03,chen-10prb155138,turner-11prb075102,fidkowski-11prb075103,schuch-11prb165139,chen-12s1604}. 
Examples of the former are the fractional quantum Hall effect\,\cite{tsui-82prl1559,laughlin83prl1395} and $\mathbb{Z}_2$ spin liquids\,\cite{read-91prl1773}, while those of the latter are spin-1 Haldane chains\,\cite{haldane83prl1153}, time-reversal invariant topological insulators and topological superconductors\,\cite{hasan-10rmp3045,qi-11rmp1057,Bernevig13}. For intrinsic topological order, it is well-known that local perturbations are irrelevant and cannot change the topological phase; this insight is at the heart of topological quantum computing\,\cite{nayak-08rmp1083}. Also for symmetry protected topological phases (SPTP), local perturbations are irrelevant as long as the protecting symmetries are not broken. One can easily convince oneself that  {\it global} breaking of, say,  time-reversal symmetry destroys the topological insulator phase. It is, however, less clear what the effect of a {\it local} perturbation is when it breaks the protecting symmetry. From a fundamental perspective, it is important to understand whether or not a SPTP becomes fragile only because the protecting symmetry is broken locally on a single or a few sites in a macroscopic sample consisting of thousands or millions of sites. From a technological aspect, this question becomes interesting as well: Is it possible to change the topological phase or the systems character  by manipulating a small region of the system?

In this paper, we aim to shed some light on this fundamental questions by considering a simple example of SPTP in the two-dimensional topological insulator candidate materials {\it silicene}, {\it germanene}, and {\it stanene}.
At a boundary between two topologically different phases (including the vacuum), metallic edge modes must appear and traverse the bulk gap. We investigate the effect of various perturbations applied (i) to one half of the system, (ii) to the lattice sites belonging to the edges, and (iii) to isolated {\it islands} consisting of a few edge-sites only.
It is intriguing that we can realize quantum spin Hall (QSH) effect without edge states.
Our analysis paves the way to some very interesting applications which base on local edge manipulations. We propose that silicene, germanene, and stanene nanoribbons with {\it manipulated edges} can be used for giant magneto-resistance and as a perfect spin-filter.

\begin{figure*}
\centering
\includegraphics[width=\linewidth]{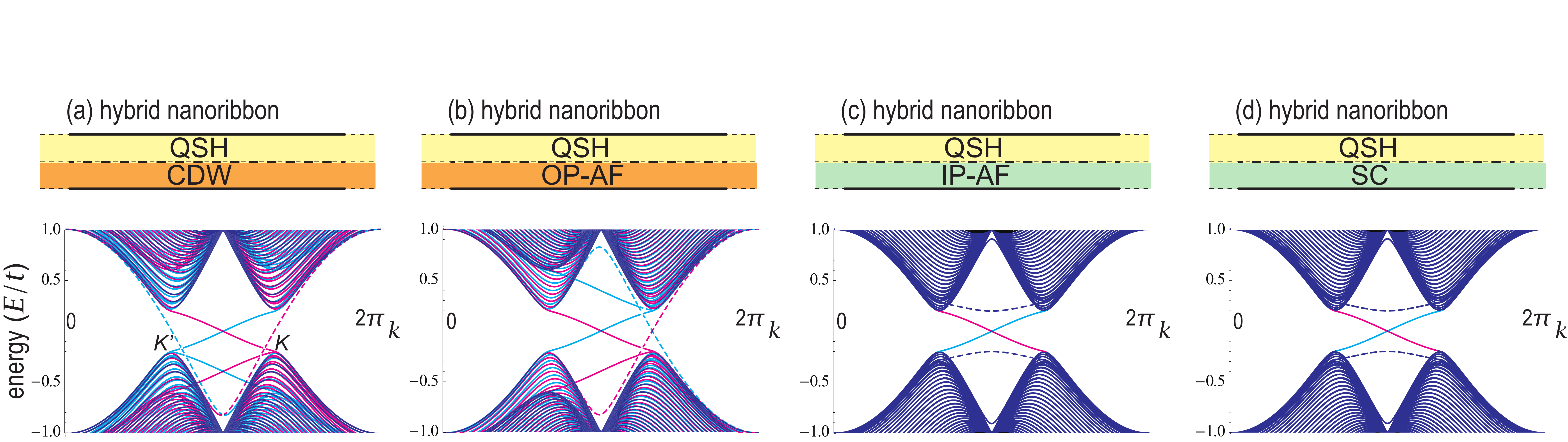}
\caption{(Color online). QSH nanoribbons (with $\lambda=0.2\,t$ and a ribbon width $W=128$ atoms) where one half of the system is exposed to one of the following perturbations. 
(a) CDW order leading to inner edge states at the boundary between the CDW and QSH regions. We have set $M=2\lambda$.
(b) OP-AF order leading to inner edge states at the boundary between the magnetic and QSH regions. We have set $m_{\rm OP}=2\lambda$.
(c) IP-AF order not leading to inner edge states. We have set $m_{\rm IP}=\lambda$.
(d) Superconducting order not leading to inner edge states. 
Magenta (cyan) solid [dotted] lines represent up-spin (down-spin) polarized modes localized at the 
outer [inner] edge. We have set $\Delta =\lambda$.
}
\label{fig:half-hybrid-ribbon}
\end{figure*}

\section{Quantum spin Hall effect in silicene, germanene, and stanene}
Silicene, germanene, and stanene are candidate materials for two-dimensional time-reversal invariant topological insulators\,\cite{liu-11prl076802,liu-11prb195430,xu-13prl136804}.
Under symmetry-breaking external fields\,\cite{ezawa12prl055502}
even more topological phases might be realized.

Silicene is a monolayer of silicon atoms forming a buckled honeycomb lattice. 
The spin-orbit (SO) coupling is expected\,\cite{liu-11prb195430} to be 
$\lambda= 3.9$\,meV$=2.4\times 10^{-3}\,t$ with the hopping parameter $t=1.6$\,eV. 
The Rashba SO terms are present but they are negligibly small to affect our analysis\,\cite{liu-11prb195430,ezawa13arXiv:1310.3536}. 
Germanene is a single layer of germanium atoms forming a buckled honeycomb lattice. The SO coupling is about ten times larger than that of silicene\cite{liu-11prb195430}, 
$\lambda=43$\,meV$=3.3\times 10^{-2}\,t$ with $t=1.3$\,eV.
Stanene is the tin-version of silicene and germanene: a single layer of tin atoms forming a buckled honeycomb lattice. Recent {\it ab initio} calculations have revealed\cite{xu-13prl136804} that 
$\lambda=0.1$\,eV$=0.077\,t$ with $t=1.3$\,eV.
Furthermore, it could be that
$\lambda=0.3$\,eV$=0.15\,t$ with $t=2$\,eV in fluorinated stanene.
A huge bulk gap due to the SO coupling would make it possible to materialize a topological insulator 
at room temperature.
We take $\lambda=0.2\,t$ for illustrations in what follows.

Both silicene, germanene, and stanene are well described by Kane and Mele's minimal topological insulator model on the honeycomb lattice\,\cite{kane-05prl146802,kane-05prl226801,liu-11prb195430},
\begin{equation}\label{ham-km}
\mathcal{H}_{\rm KM} = -t \sum_{\langle ij \rangle \alpha} c_{i\alpha}^\dag c_{j\alpha}^\pd 
+ i \frac{\lambda}{3\sqrt{3}} 
\sum_{\langle\!\langle ij \rangle\!\rangle \alpha\beta} 
\nu_{ij} c_{i\alpha}^\dag \sigma^z_{\alpha\beta} c_{j\beta}^\pd , \\[5pt]
\end{equation}
where $\sigma^z$ is the Pauli matrix associated with spin degree of freedom.
The first term $\propto t$ represents the nearest-neighbor hopping resulting in a Dirac semi-metal.  The second term $\propto\lambda$ represents the intrinsic SO term $L_z S_z$ corresponding to an imaginary spin-dependent second-neighbor hopping in real space, where
$\nu_{ij}=+1~(-1)$ if the next-nearest-neighboring hopping is anti-clockwise (clockwise) with respect to the positive $z$ axis. 
Note that since only $\sigma^z$ is involved, the spin component $S_z$ is a good quantum number. 

The SO coupling $\lambda$ determines for Hamiltonian \eqref{ham-km} the bulk gap, $\eps_{\rm gap} = 2\lambda$. Changing the value of $\lambda$ does not change any of our results. To guarantee stability of the topological phases, for very small values of $\lambda$ the ribbon width $W$ must be sufficiently large making numerical simulations efficient. But the qualitative results are not affected when varying $\lambda$. Therefore we consider throughout the paper the SO coupling $\lambda=0.2$\,eV, hopping amplitude $t=1$\,eV, and a ribbon width of $W=128$ sites. We emphasize that all results apply, hence, not only to stanene but also to silicene and germanene.

In the following, we consider various perturbation terms which lead to topological phase transitions from the QSH phase into the trivial phase. First, we consider the charge-density wave (CDW) term (staggered potential) which breaks the spatial inversion symmetry: 
\begin{equation}\label{pert-sem}
\mathcal{H}_{\rm CDW} = M \sum_{i\alpha} \left( a_{i\alpha}^\dag a_{i\alpha}^\pd - b_{i\alpha}^\dag b_{i\alpha}^\pd \right),
\end{equation}
where the annihilation operator $a_{i\sigma}$ ($b_{i\sigma}$) acts on sublattice $A$ ($B$). Second, we consider the superconducting $s$-wave pairing term which breaks the U(1) particle conservation:
\begin{equation}
\mathcal{H}_{\rm SC} = \sum_i \left( \Delta c_{i\up}^\dag c_{i\dw}^\dag + \Delta^\ast c_{i\dw}^\pd c_{i\up}^\pd \right). 
\end{equation}
We also consider two magnetic terms, an out-of-plane antiferromagnetic (OP-AF) exchange field (\ie the magnetization is perpendicular to the plane),
\begin{equation}
\mathcal{H}_{\rm OP} = m_{\rm OP} \sum_{i\alpha\beta} 
\left( a_{i\alpha}^\dag \sigma^z_{\alpha\beta} a_{i\beta}^\pd -   b_{i\alpha}^\dag \sigma^z_{\alpha\beta} b_{i\beta}^\pd\right),
\end{equation}
and an in-plane antiferromagnetic (IP-AF) exchange field (\ie the magnetization lies in the plane),
\begin{equation}\label{pert-mip}
\mathcal{H}_{\rm IP} = m_{\rm IP} \sum_{i\alpha\beta} \left( a_{i\alpha}^\dag \sigma^x_{\alpha\beta} a_{i\beta}^\pd -   b_{i\alpha}^\dag \sigma^x_{\alpha\beta} b_{i\beta}^\pd\right).
\end{equation}
The latter two terms break the time-reversal symmetry. The IP-AF exchange field does also break the $S_z$ spin symmetry, while the OP-AF exchange field preserves it. Note that the IP-AF exchange field corresponds to the mean-field description\,\cite{rachel-10prb075106,wu-12prb205102,reuther-12prb155127}
of the correlated extension of Hamiltonian \eqref{ham-km}.
Any of these perturbations may turn the system into the trivial phase.
Note that there are several distinguishable states belonging to the trivial phase which, in principle, might all be adiabatically connected with each other.

\section{Hybrid nanoribbons}
The QSH phase as present in \eqref{ham-km} is protected by the U(1) particle conservation, 
time-reversal, and $S_z$ spin symmetries.
If these symmetries are globally broken, the edge states become gapped immediately and the phase can be adiabatically connected to a conventional trivial band-insulator phase. But as long as these symmetries are intact, there must be metallic edge states at the boundaries between the topological phase and any topologically trivial phase (including the vacuum).
We investigate the effect of the aforementioned perturbations applied to only one half of the nanoribbon\,\cite{ezawa13prb161406}.
That is, one half of the ribbon is described by Eq.\,\eqref{ham-km} and the other half by the Hamiltonian\,\eqref{ham-km} plus one of the additional terms \eqref{pert-sem}--\eqref{pert-mip}. We call it a {\it hybrid nanoribbon}.
It turns out that such a nanoribbon is separated into two topologically different regions with a phase boundary. Between the two topologically different regions an inner phase boundary or an inner edge is present.

We show the band structures of hybrid nanoribbons in Fig.\,\ref{fig:half-hybrid-ribbon},
where various perturbations are applied to only one half of the nanoribbon. 
The helical edge modes at the upper edge remain unchanged, since they are not exposed to the applied field. 
In contrast, the helical edge modes at the lower edge disappear because this region of the nanoribbon has changed into a topologically trivial phase due to the applied field. 
An important question is whether inner edge modes emerge. It turns out that this depends on the type of applied field or perturbation and we have to distinguish between two classes of such perturbations.

\begin{figure}[t!]
\centering
\includegraphics[width=0.47\textwidth]{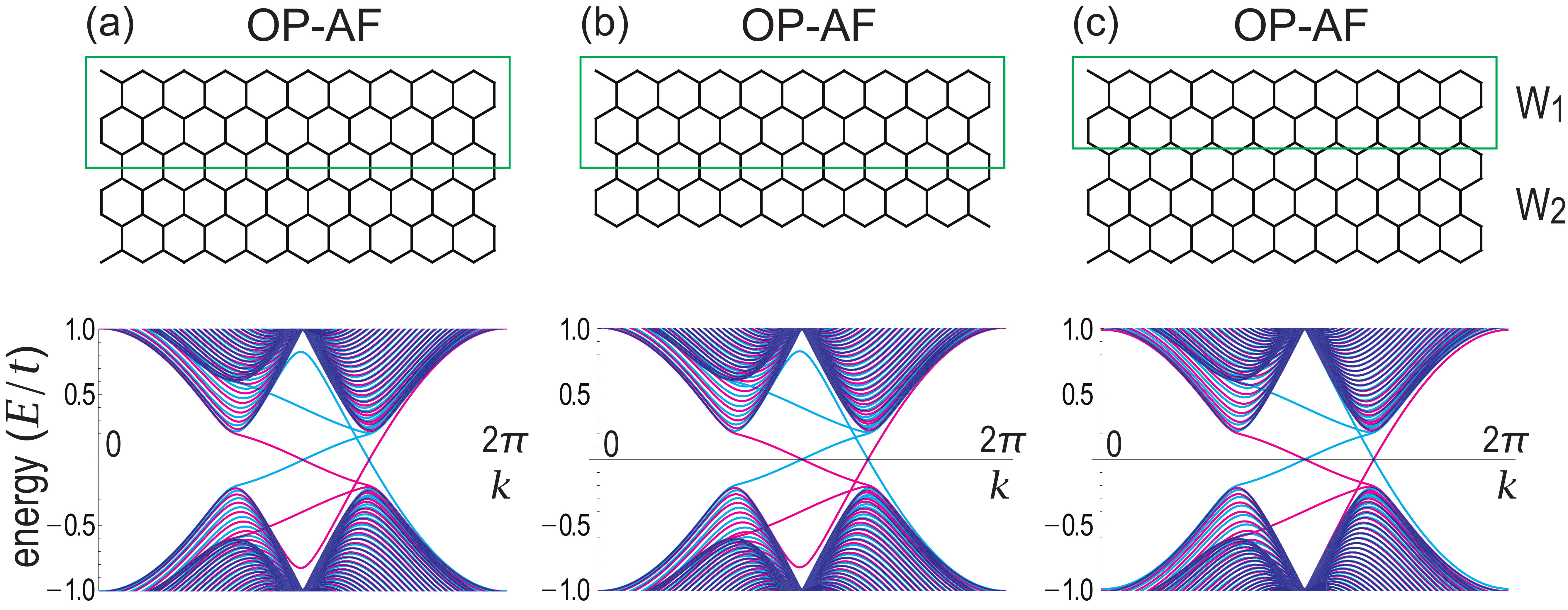}
\caption{(Color online).  Band structure of ``asymmetric'' hybrid ribbons, \ie a QSH ribbon where region 1 with width $W_1$ is exposed to the OP-AF exchange field ($m_{\rm OP} = 2\lambda$) while region 2 with width $W_2$ is unaffected ($m_{\rm OP} = 0$). The total width of the ribbon is $W=W_1+W_2$.
%
(a) Nanoribbon with $W=128$ atoms where $W_1=64$ and $W_2=64$.
(b) Nanoribbon with $W=126$ atoms where $W_1=64$ and $W_2=62$.
(c) Nanoribbon with $W=128$ atoms where $W_1=63$ and $W_2=65$.}
\label{fig:InnerEdge}
\end{figure}

The first class contains the CDW and OP-AF orders, where the inner edge modes emerge 
as shown in Fig.\,\ref{fig:half-hybrid-ribbon}\,(a) and (b). 
For hybrid nanoribbons of the first class, the spin-Chern number remains as a good topological invariant. Hence the gap must collapse at the boundary and this is accomplished by the helical edge states traversing the bulk gap. The spin-Chern number becomes ill-defined and changes its value at this point.
The Fermi momentum of the original helical edge modes is present at $k=\pi/a$. 
On the other hand, the Fermi momentum of the inner edge modes resides at the $K$ or $K'$ point and it is well described by the Jackiw-Rebbi solution\,\cite{ezawa12njp033003}.

The second class contains the IP-AF and superconducting (SC) orders, where the inner edge modes do not emerge 
as shown in Fig.\,\ref{fig:half-hybrid-ribbon}\,(c) and (d). 
It has been clarified\,\cite{ezawa-13sr2790,rachel13arXiv:1310.3159,ezawa13arXiv:1310.3536} that the IP-AF and SC orders can be connected to the QSH insulator without gap closing. We conclude that the emergence of the inner edge modes has one-to-one correspondence with whether the two adjacent topological phases can be connected due to a parameter-driven quantum phase transition without gap closing\,\cite{ezawa-13sr2790,rachel13arXiv:1310.3159,ezawa13arXiv:1310.3536}.

A comment is in order. So far we have assumed that the inner boundary is located precisely at the middle of the nanoribbon.
Our results do not change even if we alter the position of the inner boundary.
For demonstration, we show the band structure of a hybrid nanoribbon in Fig.\,\ref{fig:InnerEdge}\,(b) where the width of the unperturbed region is slightly shorter leading to an asymmetric hybrid ribbon.
The band structure remains effectively unchanged compared to the symmetric case, Fig.\,\ref{fig:InnerEdge}\,(a).
As the penetration length of the edge states is as short as one site, the position of the inner boundary does not matter at all.
Moreover, we show the band structure of a hybrid nanoribbon in Fig.\,\ref{fig:InnerEdge}\,(c) where the inner boundary is shifted by one site compared to Fig.\,\ref{fig:InnerEdge}\,(a). Although the high-energy structure is slightly modified, the low-energy spectrum is unaffected showing that  the details of the boundaries are irrelevant.

\begin{figure}[t!]
\centering
\includegraphics[width=0.46\textwidth]{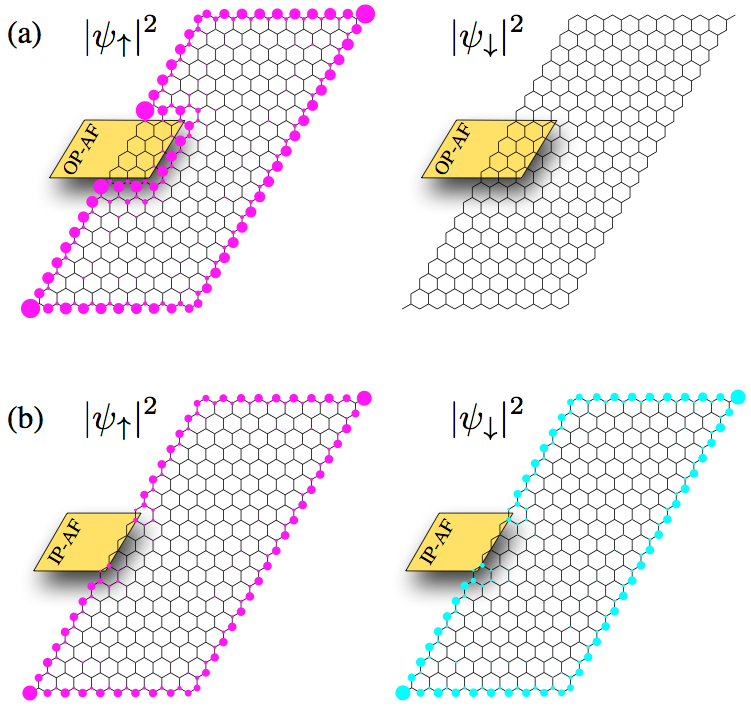}
\caption{(Color online). Local manipulations on honeycomb nanodisks ($\lambda=0.5\,t$) consisting of 800 lattice sites: the local density $|\psi_\sigma(x,y)|^2$ of the edge state is shown which is proportional to the radius of the dots. On the left side, $|\psi_\up|^2$ is plotted in magenta, on the right side, $|\psi_\dw|^2$ is plotted in cyan.
 (a) OP-AF exchange field ($m_{\rm OP}=2\,t$) is applied to a small region at the edge. The helical edge modes detour the perturbation. 
Note that the edge state $\psi$ corresponding to the energy closest to zero is shown; it is a pure $\up$-spin state.
(b) IP-AF exchange field ($m_{\rm IP}=2\,t$) is applied to five adjacent sites at the edge. At the perturbation, no edge state is present. Instead, the spin is flipped and sent back. 
Again the edge state $\psi$ with energy closest to zero is shown; the state has equal weights for $\up$- and $\dw$-spin components.
}
\label{fig:discs}
\end{figure}

Another important question about  the edge modes is whether they are spin polarized  when perturbations break the $S_z$ spin symmetry.
We have numerically checked that all gapless edge modes are perfectly spin polarized for all situations discussed in this paper, with exception of the scenario shown in Fig.\,\ref{fig:discs}\,(b). Note that pure $\up$-spin ($\dw$-spin) polarized edge states are always shown in magenta (cyan) throughout the paper; in case that both spin-components are involved the colors are superimposed yielding dark-blue.

\section{Edge manipulation of nanoribbons}
We have so far exposed one half of a nanoribbon to various fields (\ie perturbations) leading to topologically trivial phases in the exposed region. But what happens if we shrink the exposed region to a few sites so that one can hardly talk about a phase anymore?

\begin{figure}[t!]
\centering
\includegraphics[width=0.46\textwidth]{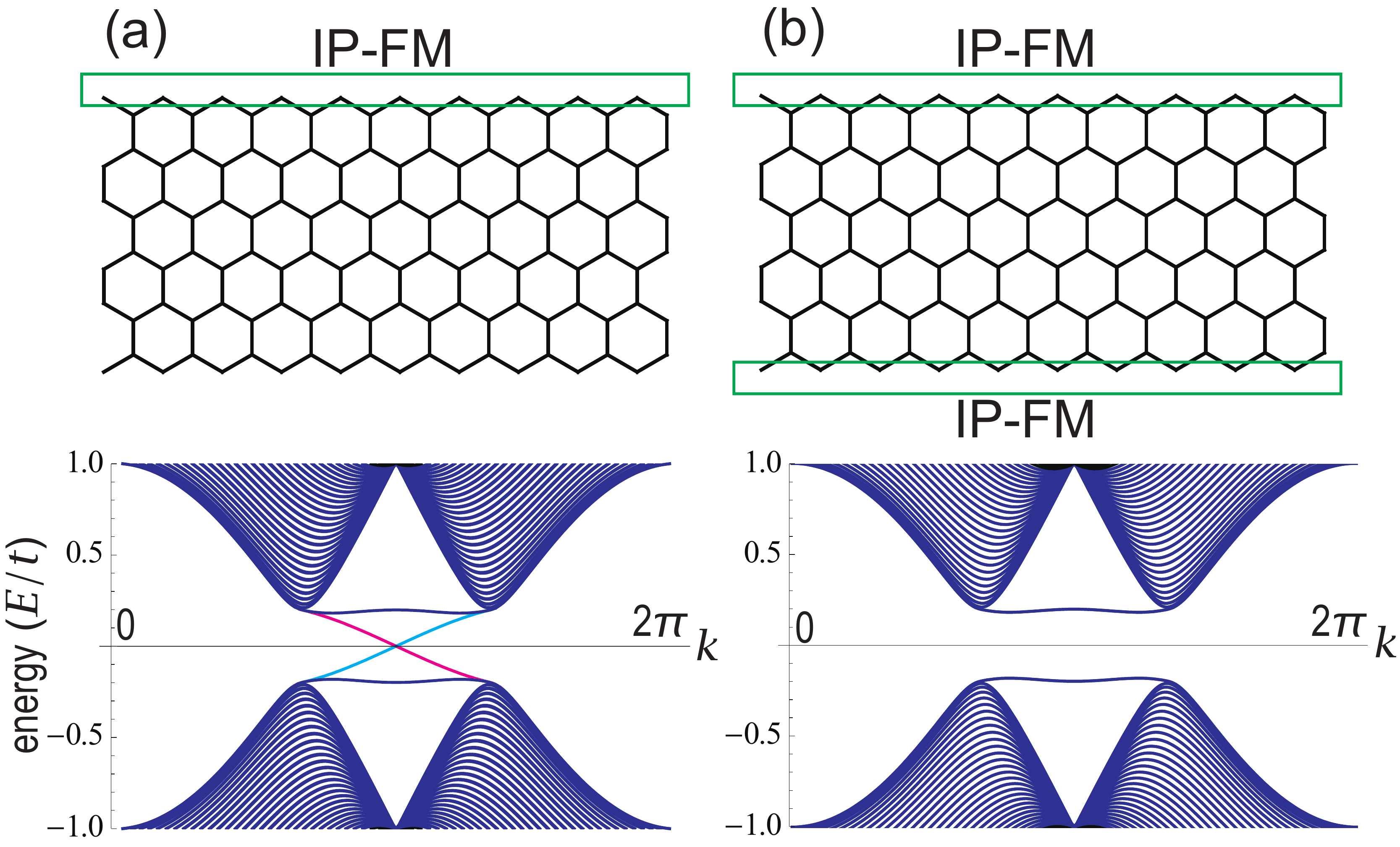}
\caption{(Color online). Nanoribbon with a width of $W=128$ atoms and SO coupling $\lambda =0.2t$.
(a) IP-FM exchange field is applied only to the outermost sites of the upper edge. Only the upper edge modes are gapped. 
(b) IP-FM exchange field is applied to the outermost sites of both edges. All edge modes are gapped.
We have set $m_{\rm IP} = \lambda$.}
\label{fig:ribbon_1edge}
\end{figure}

To answer this question, we investigate a nanoribbon where the perturbations are applied to the outermost sites of (i) a single edge or (ii) both edges: see the top panels in Fig.\,\ref{fig:ribbon_1edge}. This represents the least perturbation one can apply to a nanoribbon. 
It is worth mentioning that the antiferromagnetic (AF) and the ferromagnetic (FM) order become indistinguishable when they are applied only to the outermost sites of zigzag edges\,\cite{soriano-10prb161302}. 
For practical purposes it is easier to apply the FM field rather than the AF field. 

In addition, we have two important motivations for considering this setup: (i) it will be experimentally relevant to perform the manipulations only on the edge of the sample since the whole sample might not always be accessible due to substrate, gates etc.; (ii) in case of a large honeycomb sheet, we can safely assume that the topological phase will not be destroyed if only a small fraction of the sheet (namely the edge) is exposed to a perturbation, hence the edge manipulation should be irrelevant for the topological properties of the bulk.
In the following, we will restrict our discussion to consider the magnetic perturbations (the OP-FM field as an example of the first class and the IP-FM  field as an example of the second class). Almost identical results are derived for the cases of the CDW order and the superconductor.

By applying the IP-FM order to a single edge of the nanoribbon, the edge modes located at the same edge become immediately gapped. 
By exposing both edges of the nanoribbon to the IP-FM field, all the edge modes become gapped. Both scenarios are illustrated in Fig.\,\ref{fig:ribbon_1edge}. 
This is because the penetration length of edge modes located at zigzag edges is almost as short as a single lattice spacing.

A nanoribbon can be very wide, consisting of thousands of unit cells. It is very unlikely that the QSH phase is destroyed deep in the bulk when performing these manipulations at the edge. Nevertheless the edge states are gapped. This situation corresponds to a {\it QSH phase without edge states}. 

In the case of an armchair nanoribbon, the helical edge modes become also gapped in the presence of the IP-AF order. The magnitude of gap opening becomes larger as the region of the applied IP-AF field becomes larger. This is due to the fact that the penetration length 
is longer 
compared to zigzag edge-modes.

\section{Local manipulations of nanodisks}
We have so far considered merely perturbations affecting the whole edge. 
We proceed to investigate how a perturbation, which has the size of a few sites located at the edge, affects the QSH phase. 
We have checked various disk sizes 
ranging from 400 to 1600 lattice sites 
in order to rule out finite size artifacts.

First, the CDW and the OP-AF exchange fields create an inner boundary with helical edge modes [Fig.\ref{fig:half-hybrid-ribbon}(a) and (b)]. We can conjecture that also for a disk geometry such a perturbation will  {\it detour} the helical edge states around the exposed sites. Indeed this is what we observe for nanodisks: see Fig.\,\ref{fig:discs}(a) for an example with applied OP-AF field. The perturbation is applied in the yellow region. The shown edge state density $|\psi|^2$ corresponds to a pure $\up$-spin state with energy 
$E=-5.560\times 10^{-3}\,t$. Note that this state is degenerate with $\tilde\psi$ which is a pure $\dw$-spin state.

Second, the exposed edge modes disappear in the nanoribbon for the applied SC and IP-AF orders [Fig.\ref{fig:half-hybrid-ribbon}(c) and (d)]. 
Does such a perturbation detour the edge states as well in case of the nanodisk? Or does it make them disappear? Neither of these guesses is correct. At the exposed sites, the edge modes disappear, but everywhere else they persist. This is accomplished by flipping the spin of the edge mode at the perturbation and send the edge mode back with reversed spin orientation. That is, the former right-moving $\up$-spin edge state is now a superposition of right-moving $\up$-spin (magenta dots) and left-moving $\dw$-spin (cyan dots) edge state. For an illustration of the $\up$- and $\dw$-spin densities see Fig.\,\ref{fig:discs}\,(b), where the yellow region indicates the lattice sites which are exposed to the IP-AF field. The shown edge state density $|\psi|^2$ corresponds to the state with energy 
$E=-1.617\times 10^{-2}\,t$. Note that there is no fundamental reason to chose such a large SO coupling as in Fig.\,\ref{fig:discs}. In order to localize the edge states {\it at the edges} for a small disk size, large $\lambda$ is required. By considering much larger disks we could have used $\lambda=0.2\,t$ instead as in the other figures.

\begin{figure}[t]
\centering
\includegraphics[width=0.50\textwidth]{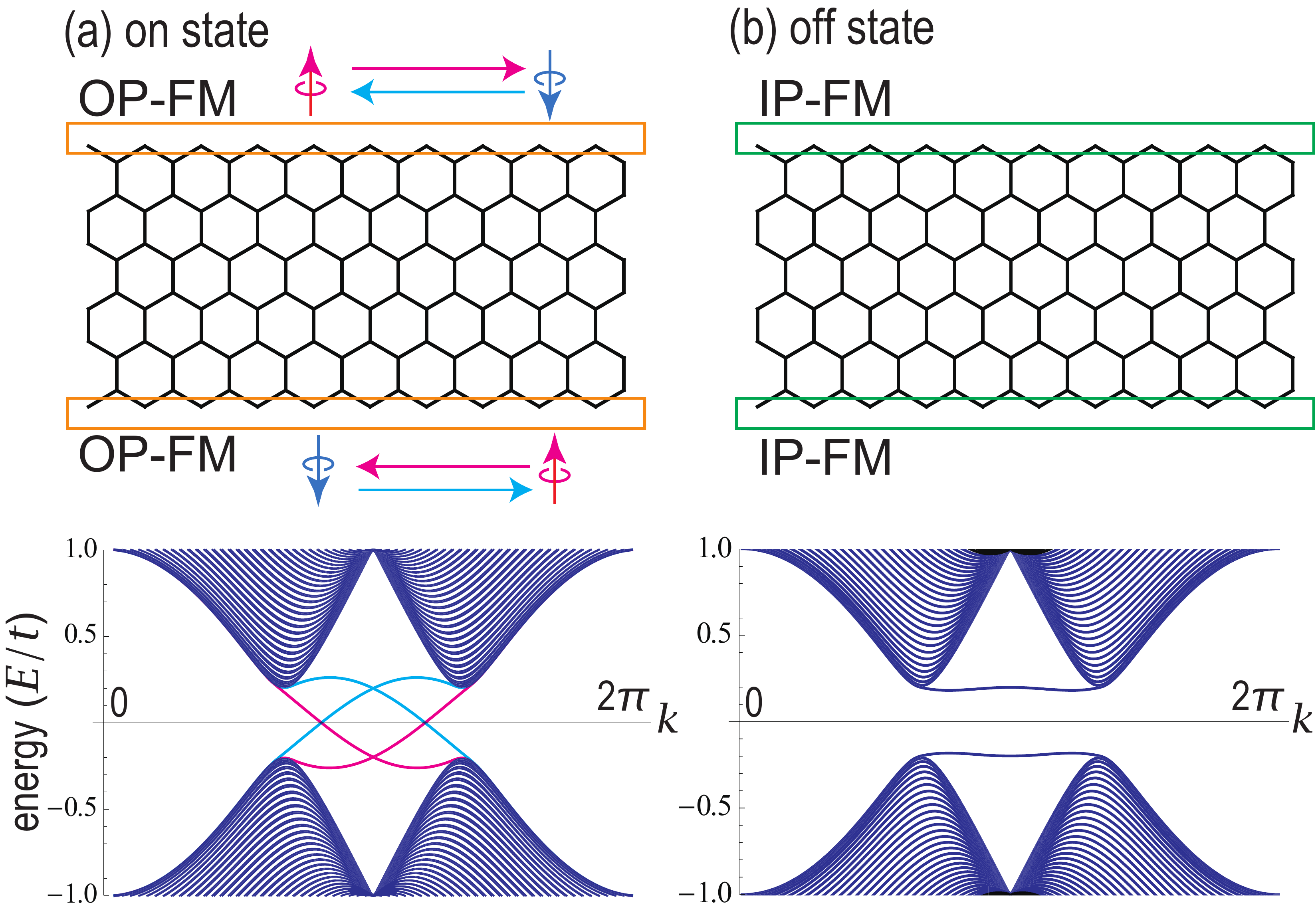}
\caption{(Color online). Illustration of a giant magneto-resistance.
We take a nanoribbon ($W=128$), to which the FM order is applied only at the outermost sites of both edges. We have set $\lambda =0.2t$.
(a)~When the FM order is out-of-plane ($\theta=0$), there are edge states yielding a quantized conductance. We have set $m_{\rm OP} = \lambda$.
(b)~When the FM order has finite in-plane component ($\theta>0$), there are no metallic edge modes. We have set $m_{\rm IP} = \lambda$.
}
\label{fig:GiantMR}
\end{figure}

\section{Applications}
The previous considerations and results can be directly used to propose some applications.

\subsection{Giant magneto-resistance}
The first application is a giant magneto-resistance. We consider the following setup (see Fig.\,\ref{fig:GiantMR}): a nanoribbon (or equivalently a two-dimensional plane where we assume one direction to be infinitely extended) whose edges are exposed to a magnetic field with variable direction of magnetization. In the simplest case, we assume that the direction can be adjusted in the $xz$ plane by parametrizing
\begin{equation}
\bs{m}_{\rm edge} = m \, \big(\, \sin{\theta}, 0,  \cos{\theta}\,\big)\ .
\end{equation} 
We assume that the magnetization strength is sufficiently large, $m \sim \lambda$, and $\theta$ can be controlled externally. The edge magnetization is chosen such that it corresponds to an OP-AF (IP-AF) exchange field for the limiting case $\theta=0$ ($\theta=\pi$). For $0<\theta<\pi$ both OP-AF and IP-AF components are present.

For $\theta=0$, there are edge states which contribute to a quantized spin current, see for the corresponding edge states the left bottom panel in Fig.\,\ref{fig:GiantMR}. This corresponds to the ``on state''. A tiny change in $\theta$ causes an in-plane magnetic contribution and the edge modes acquire a gap. By adjusting the chemical potential at zero energy, a tiny gap in the edge modes causes vanishing of the spin current. This corresponds to the ``off state''.
Turning $\theta$ back to zero, the conductance jumps again on a finite, quantized value. This is a giant magneto-resistance since the finite conductance jump is controlled or induced by a tiny angle of the external magnetic field. 

One might call this setup also a topological quantum transistor\,\cite{ezawa13apl172103} since the conductance can be switched by the external field.

%
\vspace{80pt}

\subsection{Perfect spin-filter}
\begin{figure}[t!]
\centering
\vspace{20pt}
\includegraphics[width=0.50\textwidth]{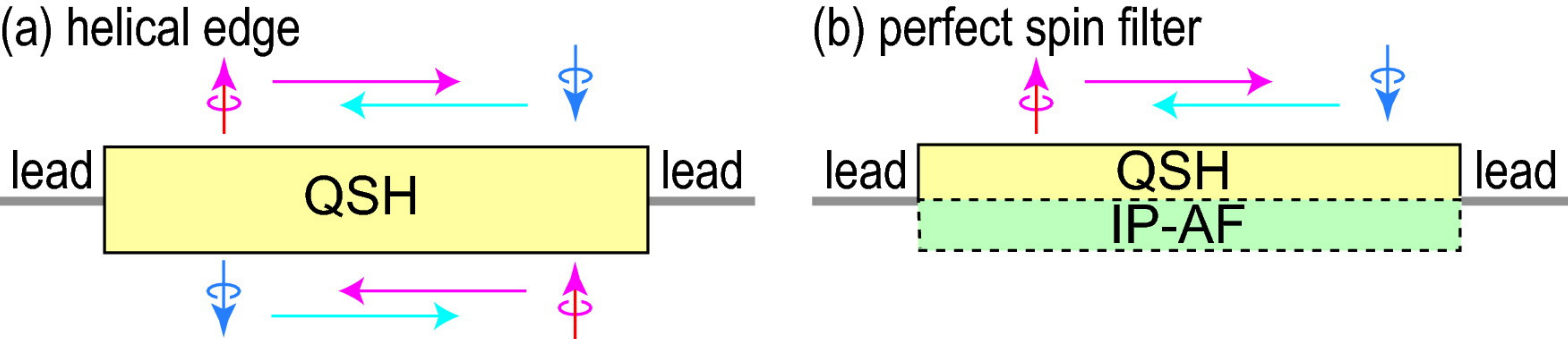}
\caption{(Color online). Illustration of a perfect spin-filter. 
A solid (dotted) boundary represents an edge where metallic edge modes (do not) emerge.
(a) In general, helical modes circulate around a sample.
(b) By introducing the IP-AF order to the lower half of the sample, helical modes only propagate along the upper edge. In this example,  only $\up$-spins are transported from the left to the right lead.
}
\label{fig:spinfilter}
\end{figure}
The second application is a perfect spin filter, which is realized when we turn on the IP-AF order only on one half of  the nanoribbon [Fig.\,\ref{fig:spinfilter}(b)]. Since helical edge states are present  
only on the other half of the nanoribbon, we have a {\it one-way helical edge state}. That is, by sending a spin-unpolarized current through the nanoribbon,  only $\up$-spins (or $\dw$-spins) can pass the nanoribbon, hence it is a spin-filter. The spin filter is {\it perfect} since the spin-momentum locking is an inescapable property of the topological insulator.
We note that usual helical edge modes circulate around the sample, that is, the direction of two helical edge modes are opposite on opposite sides of the nanoribbon [Fig.\,\ref{fig:spinfilter}\,(a)]. In the latter case, there is no spin-filtering effect. Note that similar ideas about blocking a helical edge channel have been considered in Refs.\,\onlinecite{qi-08np273,timm-12prb155456}.


\section{Discussion}
We have repeated all calculations, shown in this paper for silicene, germanene, and stanene also for the BHZ model\,\cite{bernevig-06s1757} which describes the topological insulator phase in HgTe/CdTe quantum wells\,\cite{koenig-07s766}. All our considerations about edge manipulations remain unchanged and, hence, the HgTe/CdTe quantum wells provide an equally well suited platform for a giant magneto-resistance and a perfect spin-filter.

\phantom{x}

\phantom{x}

\begin{acknowledgements}
The authors acknowledge  discussions with L.\ Molenkamp, N.\ Nagaosa, R.\ Thomale and M.\ Vojta. SR is supported by the DFG through FOR 960, the DFG priority program SPP 1666 ``Topological Insulators'' and by the Helmholtz association through VI-521. ME is supported in part by Grants-in-Aid for Scientific Research from the Ministry of Education, Science, Sports and Culture No. 25400317.
\end{acknowledgements}

\bibliographystyle{prsty}
\bibliography{hybrid-qsh} 
\end{document}